\documentclass[%
reprint,
superscriptaddress,
 amsmath,amssymb,
 aps,
showpacs,
]{revtex4-1}
\usepackage{graphicx}
\usepackage{braket}
\usepackage{bm}
\usepackage{amsmath}

\def\vect#1{\mbox{\boldmath $#1$}}

\newcommand{\bor}{${^{11}{\rm B}}$}
\newcommand{\cc}{${^{12}{\rm C}}$}
\makeindex            
 
\begin{document}
\title{2$\alpha$+$t$ cluster structure in  \bor }
\author{Bo Zhou}
\affiliation{Institute for International Collaboration, Hokkaido University, Sapporo 060-0815, Japan}
\affiliation{Department of Physics, Hokkaido University, 060-0810 Sapporo, Japan}
\author{Masaaki Kimura}
\affiliation{Department of Physics, Hokkaido University, 060-0810 Sapporo, Japan}
\affiliation{Reaction Nuclear Data Centre, Faculty of Science, Hokkaido University, 060-0810 Sapporo, Japan}

\date{\today}
	
\begin{abstract}
 The 2$\alpha+t$ cluster structure in $^{11}$B is investigated by the microscopic generator
 coordinate  method (GCM) with the Brink cluster wave functions. With a proper choice of the
 parameters of  the effective interaction, the calculated energy spectrum shows reasonable
 agreement with the observed low-lying spectra of both parities. On the basis of the calculated 
 radii, monopole and $B(E2)$ transition strengths, several developed cluster states of $^{11}$B
 are suggested. For the negative-parity states, in addition to the well-known $3/2_3^-$ cluster
 state, the $1/2_2^-$ and $5/2_3^-$ states are also proposed as the well-developed cluster
 states. For the  positive-parity states, it is found that many states around the 2$\alpha+t$
 threshold show the feature of developed clusters. In  particular, the $1/2_2^+$ state is found
 to have a linear-chain-like structure, which is  consistent with the previous antisymmetrized
 molecular dynamics calculation, but contradicts to  the orthogonality condition model
 calculation. It is also found that many of these positive-parity cluster candidates have the
 non-negligible isoscalar dipole transition strengths, which require the experimental
 confirmation.
\end{abstract} 
\pacs{21.60.Gx, 21.10.Ky, 27.20.+n}
\maketitle

\section{Introduction}
The study of cluster states of light nuclei \cite{wildermuth_unified_1977,fujiwara_chapter_1980,kanada-enyo_antisymmetrized_2003,
freer_clustered_2007,horiuchi_recent_2012,kimura_antisymmetrized_2016} has always been an important subject in
nuclear physics, which provides us with a new perspective for understanding nuclear structure
and many-body problem in atomic nuclei. The \cc\ is one of the most typical clustered nuclei. In
particular, its famous Hoyle state is known as an $\alpha$ condensed
state~\cite{tohsaki_alpha_2001,tohsaki_colloquium_2017} or gas-like state. This kind of well-developed cluster
states shows us a novel motion of clusters in a nucleus, which cannot be explained by the
single-particle picture. In this decade, the search for the gas-like or well-developed cluster
states~\cite{von_oertzen_nuclear_2006,itoh_candidate_2011,yang_observation_2014,he_giant_2014,zhou_breathing-like_2016,baba_structure_2016} in self-conjugate $n\alpha$ nuclei and also in non-$n\alpha$ nuclei attract great interest
both in experiment and theory.

In analogy to the 3$\alpha$ cluster structure in \cc, it is a very interesting subject to
investigate the developed 2$\alpha+t$ cluster structure around the threshold energy in the 
non-self-conjugate nucleus \bor. In particular, it is the central interest in this study if we
can find the  gas-like cluster state analog to the Hoyle state in $^{12}{\rm C}$.  
The early orthogonality condition model (OCM) calculation by Nishioka {\it et
al.}~\cite{nishioka_structure_1979} showed that the $3/2^-_3$ state at 8.56 MeV and the several negative-parity states around 10 MeV are the promising candidates for the well-developed $2\alpha+t$ clustering, while the low-lying states are the compact shell-model-like states.  They also concluded that the positive-parity states have the transient nature between the shell and cluster structure in \bor. 

Decades later, the antisymmetrized molecular dynamics (AMD) calculations showed that~\cite{suhara_cluster_2012} the
$3/2^-_3$ state has the prominent $2\alpha+t$ cluster structure without any a-priori assumption on
the cluster formation.  They also showed that~\cite{suhara_cluster_2012} the $3/2^-_3$ state has the strong IS monopole
transition strength from the ground state because of its spatially extended cluster structure
whose matter radius is 2.65 fm. This enhancement of the IS monopole transition strength is in good
accordance with the observation~\cite{kawabata_cluster_2007}. On the basis of the analysis using the Brink wave function, they 
showed that~\cite{,kanada-enyo_2alpha+triton_2015} the $t$ cluster has broad spatial distribution around the  $2\alpha$ cluster
core, which is the reason why they concluded the $3/2^-_3$ state as the Hoyle analog state.

More recently, Yamada {\it et al.}~\cite{yamada_++t_2010} renewed the Nishioka's work by performing the OCM calculation with much larger model space. They confirmed that the $3/2_3^-$
state has a $2\alpha+t$ cluster structure with a large matter radius of 3.00 fm. Furthermore, they reported that the $1/2_2^+$ state at 12.56 MeV has a dilute cluster structure with a very large
nuclear radius of 6 fm.
By analyzing the cluster orbits occupied by the $\alpha$ and $t$ clusters, they concluded that the $1/2_2^+$  state should be regarded as the  Hoyle analog state as all clusters are approximately occupying $s$-wave states, while they are not in the $3/2^-_3$ state.   
Recently, by a new measurement of $\alpha$ resonant scattering on $^{7}$Li, Yamaguchi {\it et al.}~\cite{yamaguchi__2011} did not observe the strong resonance $1/2^+$  state at 11.7$-$13.1 MeV, instead, they proposed the existence of another $3/2^+$ or $9/2^+$ state at 12.63 MeV. So far, the experimental counterpart for this  $1/2_2^+$  state  has not been identified yet.

Thus, two different Hoyle analog states, the $3/2^-_3$  and $1/2^+_2$ states, were suggested as 
the candidates of the Hoyle analog state by two different types of the theoretical models, AMD~\cite{suhara_cluster_2012,kanada-enyo_2alpha+triton_2015}
and OCM~\cite{yamada_++t_2010}. The former (the $3/2^-_3$ state) is experimentally identified well by its pronounced IS
monopole transition, but cannot have the $s$-wave nature because of its spin and parity. On the
other hand, the $1/2^+_2$ state can have the $s$-wave nature which is the most prominent
characteristics of the Hoyle state. However, its existence~\cite{yamaguchi__2011} is experimentally uncertain as it
may be a broad resonance and cannot be populated by the IS monopole transition from the ground
state. It is also noted that no microscopic cluster models report the existence of the 
$1/2^+$ state with the $s$-wave nature, and only the semi-microscopic calculations (OCM) report
it. 

This contradicting and puzzling situation motivated us to conduct the study of the $2\alpha+t$
cluster states based on the microscopic cluster model. We will study the cluster states in the
framework of the generator coordinate method (GCM) using the Brink wave functions as the basis
wave functions. Based on the excitation energies, radii, and the transition strengths, we will
discuss the possible developed cluster states in \bor.  Our main interests in this work are as
follows. (1) The property of the $3/2^-_3$ state. Is it consistent with the other theories and
experiments? (2) Are there other pronounced cluster states in the negative-parity
states other than $3/2^-_3$ state? 
(3) Is there the $1/2^+$ state with $s$-wave nature by the microscopic model calculation?
(4) What is a good experimental probe for the positive-parity cluster states ?

The paper is organized as follows. In Sec.~\ref{sec2}, we present the formulations of the GCM with 2$\alpha+t$ Brink wave functions for \bor. Then we will show the results and make discussions in Sec.~\ref{sec3}. Finally, we present the summary in Sec.~\ref{sec4}. 

\section{\label{sec2} Theoretical framework} 

\subsection{Hamiltonian and GCM wave function}

The Hamiltonian used in the present study includes the kinetic energy, effective nucleon-nucleon
and Coulomb interactions,  
\begin{align}
 H=\sum_{i=1}^{11} t_i - t_{\rm cm} + \sum_{i<j}^{11}v_{NN}+ \sum_{i<j}^{11}v_{\rm Coul},
\end{align}
where $t_{\rm cm}$ denotes the center-of-mass kinetic energy which is exactly removed from the total
energy. As the effective nucleon-nucleon interaction, we employed the Volkov No.2 interaction~\cite{volkov_equilibrium_1965} combined
with the spin-orbit part of the G3RS interaction~\cite{yamaguchi_effective_1979,okabe_structure_1979} which is given as,
\begin{align}
 v_{NN} =& \sum_{n=1}^2 v_ne^{- r_{ij}^2/a_n^2}(W+BP_\sigma-HP_\tau-MP_\sigma P_\tau)\nonumber\\
 &+\sum_{n=1}^2 w_ne^{-  r_{ij}^2/b_n^2 } P(^3O)\bm L\cdot \bm S.
\end{align}
Here $P_\sigma$, $P_\tau$, and $P(^3O)$ denote the spin and isospin exchange operators and the
projection operator to the triplet-odd states, respectively. The parameter set is listed in
Table. \ref{tab:parameter}, which is slightly modified from that adopted in Ref.~\cite{suhara_cluster_2012} to reproduce the
binding energy of the ground state and the splitting between the $3/2^-_1$ and $1/2^-_1$ states, 
simultaneously.

\begin{table}[thb]
\caption{Adopted parameter set for the Volkov No.2 interaction and the spin-orbit part of the G3RS
 interaction. The units of $v_n$ and $w_n$ are MeV. The units of $a_n$ and $b_n$ are fm and
 fm$^{-2}$, respectively }\label{tab:parameter} 
\begin{ruledtabular}
\begin{tabular}{cccccccc}
 $v_1$&$v_2$&$a_1$&$a_2$&$W$&$B$&$H$&$M$\\
 $-60.65 $&$61.14 $&$1.80 $&1.01 &0.41&0.125&0.125&0.59\\\hline
 $w_1$&$w_2$&$b_1$&$b_2$\\
 $2800 $&$-2800$ &$0.4472 $&$0.60 $
 \end{tabular}
\end{ruledtabular}
\end{table}

In this study, we employ the  Brink wave function~\cite{brink_alpha-particle_1966} for the
basis wave function of the GCM calculation. The system composed of the 2$\alpha+t$ clusters
located at $\bm R_1, \bm R_2$, and $\bm R_3$ is described as follows,   
\begin{align}
 \Phi(\{R\},s)&=
 \mathcal{A}[\Phi_\alpha(\vect{R}_1)\Phi_\alpha(\vect{R}_2)\Phi_t(\vect{R}_3,s) ],\\
 \Phi_\alpha(\bm R)&=\mathcal A\{\phi(\bm R)\chi_{n\uparrow}\cdots
 \phi(\bm R)\chi_{p\downarrow}\},\\
 \Phi_t(\bm R,s)&=\mathcal A\{\phi(\bm R)\chi_{n\uparrow}\phi(\bm R)\chi_{n\downarrow}
 \phi(\bm R)\chi_{ps}\},\\
 \phi(\vect{R})&=\Bigl(\frac{1}{\pi b^2}\Bigr)^{3/4} 
 e^{-\left(\vect{r}-\vect{R}\right)^2/(2b^2)},
\end{align}
where $\Phi_\alpha$ and $\Phi_t$ denote the wave functions of $\alpha$ and $t$ clusters,
respectively. $\phi(\vect{R})$ is the Gaussian wave packet for the single nucleon located at $\bm R$.
The size parameter $b$ is chosen as $1/(2b^2)=0.235$ fm$^{-2}$.
$\chi_{p\uparrow},...,\chi_{n\downarrow}$ represent the spin-isospin wave functions,
and the spin of the proton in the $t$ cluster denoted by $s$ is either of up or down. $\bm R_1$, 
$\bm R_2,$ and $\bm R_3$ are the generator coordinates and abbreviated as 
$\{R\}=\{ \vect{R}_1,\vect{R}_2,\vect{R}_3 \}$. 
The condition $4\bm R_1+4\bm R_2+3\bm R_3=0$ is imposed to remove the center-of-mass motion.

By the parity and angular-momentum projections, we obtain the projected wave functions,
\begin{align}
 \Phi^{J\pi}_{MK}( \{R\} ,s )= P^J_{MK}(\Omega) P^{\pi} \Phi(\{R\} ,s ),
\end{align}
where $P^\pi$ and $P^{J}_{MK}(\Omega)$ denote the parity and angular momentum projectors. 
Then the projected basis wave functions with different values of the generator coordinates 
$\{R\}$, $s$, and different projections of the angular momentum $K$ are superposed,
 \begin{align}
 \label{gcmwf}
  \Psi^{J\pi}_{M } =\sum_{\{R\}sK}  g_{\{R\}sK} \Phi^{J\pi}_{MK}( \{R\},s ).
 \end{align}
The coefficients of the superposition $g_{\{R\}sK}$ and the eigenenergies $E$ are obtained by
solving the Hill-Wheeler equation~\cite{ring_nuclear_2004}, 
\begin{eqnarray} 	
 \sum_{\{R'\}s'K'} g_{\{R'\}s'K'}
  \braket{\Phi^{J\pi}_{MK}(\{R\},s)|H|\Phi^{J\pi}_{MK'}(\{R'\},s')}\nonumber\\
 =E \sum_{\{R'\}s'K'} g_{\{R'\}s'K'}
  \braket{\Phi^{J\pi}_{MK}(\{R\},s)|\Phi^{J\pi}_{MK'}(\{R'\},s')}.
\end{eqnarray}
From the GCM wave function, we can directly calculate physical quantities such as radii and the
transition probabilities. 

In the practical calculation, the generator coordinates $\{R\}=\{\bm R_1, \bm R_2, \bm R_3\}$ are
chosen so that their inter-distances $R_{ij}=|\bm R_i-\bm R_j|$ range from 1 to 6 fm with an
interval of 1 fm. This choice of the generator coordinate, together with the degree-of-freedom
of the triton spin direction, yields 186 Brink wave functions to be used as the basis
wave function of the GCM calculation. To check the stability of the obtained GCM wave functions,
we also performed GCM calculations by adopting larger model spaces for comparison. For example, we
increased the maximum length of the inter-distance up to 8 fm, and found that the obtained
states around and below the $2\alpha+t$ threshold, which are of our interest in the present study,
are almost stable. More importantly, we confirmed that the general feature of the physical
quantities such as the IS monopole transitions are also stable. However, it is also noted that the
highly excited states lying well above the $2\alpha+t$ threshold are sensitive to the choice of
the model space because of their strong coupling with the many-body continua. This requires the 
application of some methods to describe broad resonances, which will be our main focus in the next
work.  

\subsection{Isoscalar monopole, dipole, and electric quadrupole transitions}
Recent years, the isoscalar monopole (ISM) transition has been regarded as a powerful
probe to identify the developed cluster states, since it strongly induces clustering
as proved by Yamada {\it et al.} \cite{yamada_monopole_2008,yamada_isoscalar_2012}. Indeed, many cluster states in light stable nuclei 
including the $3/2^-_3$ state of $^{11}{\rm B}$ are experimentally known to have enhanced ISM 
transition strength from the ground state. 
In this study, we utilize the ISM strengths from the low-lying $1/2^+_1$ and $5/2^+_1$ states
to identify the clustered $1/2^+$ and $5/2^+$ states, although the direct measurement of the
transition strength is not easy.
The ISM operator $\mathcal{M}^{\rm ISM}$ and the reduced transition strength from the state 
$\ket{J_iM_i}$ to the state $\ket{J_fM_f}$ are defined as,   
\begin{align}
\mathcal{M}^{\rm ISM}&=\sum_{i=1}^A (\bm r_i - \bm r_{\rm cm})^2, \\
 B(ISM;J_i  \rightarrow J_f)&=|\braket{J_fM_f|\mathcal{M}^{\rm ISM}|J_iM_i} |^2,
\end{align}
where  $\bm r_i$ and $ \bm r_{\rm cm}$ denote the $i$th nucleon coordinate and the center-of-mass,
respectively. 

In addition to the ISM transition, the isoscalar dipole (ISD) transition has been proposed as an another probe~\cite{chiba_isoscalar_2016,kanada-enyo_isovector_2016,chiba_inversion_2017} for the clustering. The transition operator and strength are defined as,
\begin{align}
\mathcal{M}^{\rm ISD}_{\mu}&=\sum_{i=1}^A (\bm r_i - \bm r_{\rm cm})^3
    Y_{1\mu}(\widehat{\bm r_i - \bm r_{\rm cm}}), \\
   B(ISD; J_i  \rightarrow J_f)
    &=\sum_{M_f\mu}|\braket{J_fM_f|\mathcal{M}^{\rm ISD}_\mu|J_iM_i} |^2.
\end{align}
It is noted that the ISD excitation from the $3/2^-$ ground state yields the excited $1/2^+$,
$3/2^+$, and $5/2^+$ states. We expect that the magnitude of the ISD transition strengths can be a
clustering measure for these spin-parity states.

The electric quadrupole transition strength is also calculated to investigate the band structure.
Its transition operator and strength are given as,
 \begin{align}
&\mathcal M_\mu^{\rm E2}=e^2\sum_{i=1}^{A}(\bm r_i - \bm r_{\rm cm})^2
 Y_{2\mu}(\widehat{\bm r_i-\bm r_{\rm cm}})  \frac{1-\tau_{iz}}{2},\\
& B(E2; J_i  \rightarrow J_f)=\sum_{M_f\mu}|\braket{J_fM_f|\mathcal{M}^{\rm E2}_\mu|J_iM_i} |^2.
 \end{align}
Here $\tau_{iz}$ is the isospin projection of the $i$th nucleon. We use the $B(E2)$ strength 
to assign the possible band structure of the excited states.

\subsection{Overlap between GCM and projected Brink wave functions}
\begin{figure}[h]
 \centering
 \includegraphics[width=0.8\hsize]{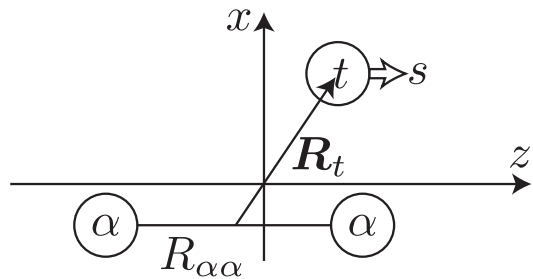}
 \caption{\label{fig_sketch} Schematic figure of 2$\alpha+t$ Brink wave function, 
 which is used to define the squared overlap given by Eq. (\ref{overlapr}).} 
\end{figure}

A GCM wave function is a superposition of many projected Brink wave functions having different
generator coordinates $\{R\}$, spin $s$, and different projections of the angular momentum $K$ in
the body-fixed frame. Therefore, it is not straightforward to assign cluster configuration to each
GCM wave function. For this purpose, it is convenient to introduce the overlap between the GCM and 
projected Brink wave functions. Here, the body-fixed frame and the Brink wave function used as a
reference state are defined as illustrated in Fig. ~\ref{fig_sketch}.
The two $\alpha$ clusters are aligned along the $z$ axis with the inter-cluster distance 
$R_{\alpha\alpha}$, and the $t$ cluster is located at $\bm R_t$ on the $zx$-plane. 
The triton spin direction is represented by $s$. 
With this Brink wave function denoted by $\Phi(R_{\alpha\alpha},\bm R_{t},s)$,
the squared overlap with the GCM wave function is defined as
\begin{align}
O(R_{\alpha\alpha},\bm R_t)&=\sum_{sKs'K'} \braket{\Psi^{J\pi}_M|
P^{J\pi}_{MK}\Phi(R_{\alpha\alpha}, \bm R_t, s)}S_{sKs'K'}^{-1}\nonumber\\
 &\times
 \braket{P^{J\pi}_{MK'}\Phi(R_{\alpha\alpha},\bm R_t, s')|\Psi^{J\pi}_M}.\label{overlapr}
\end{align}
Here $S_{sKs'K'}$ denotes the overlap between the reference Brink wave functions,
\begin{align}
S_{sKs'K'}=\braket{P^{J\pi}_{MK}\Phi(R_{\alpha\alpha},\bm R_t, s)|
    P^{J\pi}_{MK'}\Phi(R_{\alpha\alpha},\bm R_t, s')},
\end{align}
and its inverse satisfies the relationship,
\begin{align}
 \sum_{s'K'} S_{sKs'K'}S^{-1}_{s'K's''K''} = \delta_{ss''}\delta_{KK''}.
\end{align}
We expect that the magnitude of thus-defined squared overlap is a good measure to find the 2$\alpha$
clusters with inter-cluster distance $R_{\alpha\alpha}$ and the triton cluster at the position
$\bm R_{t}$. We will use this quantity to discuss the distribution of $t$ cluster around the
2$\alpha$ cluster core. It must be noted that $O(R_{\alpha\alpha},\bm R_t)$ does not
correspond to the probability, because the Brink wave functions with different cluster positions
are unorthogonal to each other. In other words, the integral of the squared overlap, 
\begin{align}
 4\pi\int dR_{\alpha\alpha}d^3R_t\ R_{\alpha\alpha}^2 O(R_{\alpha\alpha},\bm R_t),
\end{align}
is larger than unity by its definition.

\section{\label{sec3} Results and Discussions}
\subsection{Energy levels}
\begin{figure*}[t] 
\begin{center}
 \includegraphics[width=0.9\hsize]{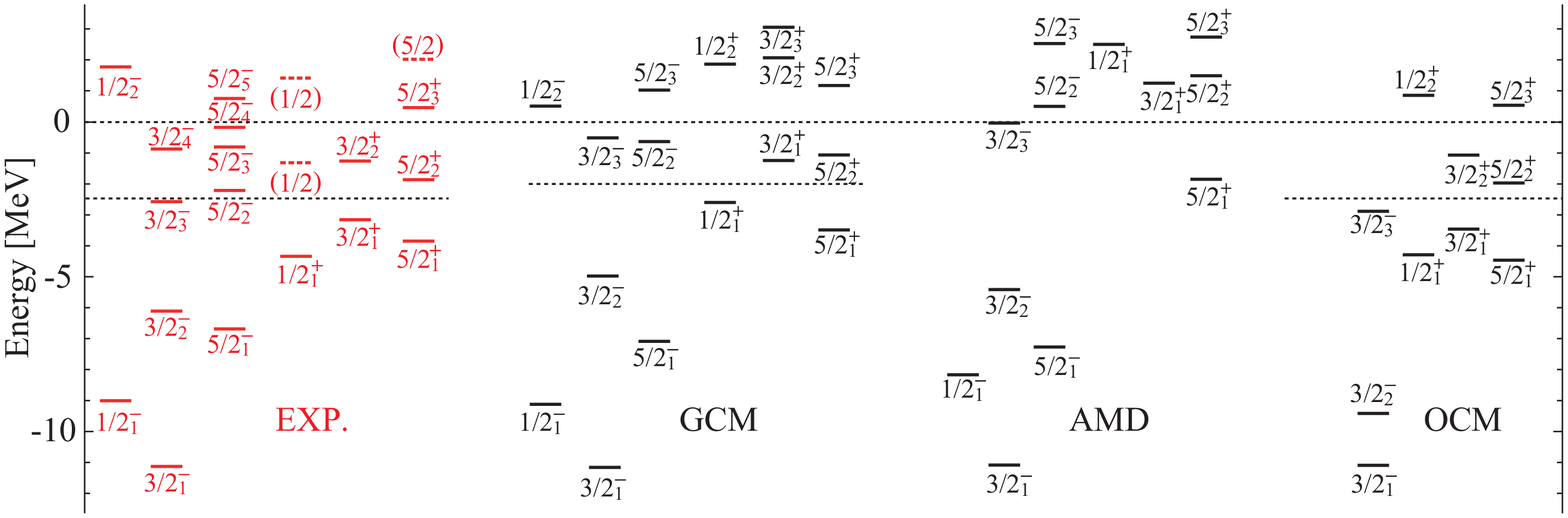}
 \caption{ \label{fig1}  Comparison of obtained energy levels from  GCM, AMD~\cite{suhara_cluster_2012}, OCM~\cite{yamada_++t_2010} calculations and experiments~\cite{kelley_energy_2012} for the \bor. The shown
 experimental and  theoretical excitation energies are relative to their corresponding
 2$\alpha$+$t$ threshold energies. The above and below dash lines represent the 2$\alpha$+$t$ and
 ${}^7$Li($3/2^-$)+$\alpha$ threshold, respectively.} 
\end{center}
\end{figure*}

Figure~\ref{fig1} shows the energy levels obtained by the present calculation (GCM) and they are compared 
with the experimental data,  OCM~\cite{yamada_++t_2010} and  AMD~\cite{suhara_cluster_2012} calculations. Note that, in the present
calculation, the  $3/2^-_1$ and $1/2^-_1$ energies are fitted to the experiment to determine the
parameter set of the effective interaction, but other states are not. We see that thus-determined
parameter set plausibly describes the low-lying positive- and negative-parity bound states
(the states below the $^{7}{\rm Li}+\alpha$ threshold) and qualitatively agrees with the
observation, although it does not bound the $3/2^+_1$ state in contradiction to the experiment.
It is also noted that the  $^{7}{\rm Li}+\alpha$ and $2\alpha+t$ threshold energies and the
energies of the  $^{7}{\rm Li} (3/2^-)$ and $^{7}{\rm Li} (1/2^-)$ are also reasonably described,
indicating that the parameter set yields reasonable inter-cluster potential for the $\alpha+t$
system.

In addition to the bound states, the calculation yields highly excited states around and above the
threshold energies. In particular, as we will discuss in the following, the states around the
$2\alpha+t$ threshold energy such as the $1/2_2^-$, $3/2_3^-$ and $1/2_2^+$ states  are the
candidates of the pronounced clustering. We see that these highly excited states can also be
assigned to the observed states from their excitation energies.  

The AMD calculation~\cite{suhara_cluster_2012} yields the negative-parity spectrum quite similar to ours, because a similar
interaction parameter set except for much weaker spin-orbit strength ($w_1=-w_2=1600$) was
employed. However, we see that it yields quite different positive-parity spectrum and no state
exists below the $2\alpha+t$ threshold. The origin of the difference between the present and AMD
calculations can be explained as follows. In the AMD study, the intrinsic wave functions were
calculated {\it before} the angular momentum projection, which does not necessarily yield the
energy minimum state {\it after} the angular momentum projection. Therefore, it can fail to
describe the lowest energy states for a given spin and parity. Indeed, as we see later, the lowest
$1/2^+$ state obtained by the AMD calculation has the linear-chain-like structure, which is
similar to our $1/2^+_2$ state. On the other hand, the $1/2^+_1$ state obtained by the present
calculation looks missing in the AMD result.

The OCM calculation~\cite{yamada_++t_2010} adopts the inter-cluster potentials which reproduce the $\alpha+\alpha$ and
$\alpha+t$ systems, and introduces phenomenological three-cluster interaction to reproduce the
binding energy of $^{11}{\rm B}$. It yields similar energy spectra to ours,
and the positive-parity spectrum shows slightly better agreement with the experiment. As explained
later, the structure of the $1/2^+$ state is quite different between the OCM and the present
calculation, which may originates in the difference of the effective interaction.  One should also
note that the observed deepest positive-parity state is the $1/2_1^+$ state, while the OCM and
present calculations fail to reproduce the correct order of the positive-parity states.

\subsection{ Negative-parity  states in \bor\ }
\subsubsection{radii and transition strengths}
We begin with the ground state of \bor\ and the excited $3/2^-$ states, namely
$J^\pi=3/2_1^-$, $3/2_2^-$, and $3/2_3^-$ states. These negative-parity states have been well 
studied by AMD, OCM, and experiments, and the comparison with these results would be useful for
verifying present calculations.

\begin{table}[!htbp]
 \centering
 \caption{\label{tb_rad} Root-mean-square (r.m.s) radii in the unit of fm for mass distribution of 
 negative-parity  states from GCM Brink, AMD~\cite{suhara_cluster_2012}, OCM~\cite{yamada_++t_2010} and experiment.} 
 \begin{ruledtabular}
 \begin{tabular}{   c   c   c   c  c  }
  State & GCM Brink& AMD & OCM & Experiment      \\	\hline
  $3/2_1^- $ & 2.38 & 2.29  &  2.22 &  2.09  $\pm$ 0.12   \\  
  $3/2_2^-$ &  2.64 & 2.46  &  2.23 &          \\   
  $3/2_3^-$ &  2.99 & 2.65  &  3.00 &        \\  
 \end{tabular}
 \end{ruledtabular}
\end{table}

In Table~\ref{tb_rad}, the root-mean-square (r.m.s) radii of the $3/2^-$ states from different
cluster models and experiment are shown. For the ground state, the  theoretical models, especially
our GCM Brink model, overestimate the experiment. This may be due to the underestimation of the
cluster distortion effect in the ground state by the cluster models. However, it is noted
that the present calculation plausibly yields the electric quadrupole moment of the ground state
4.07 $e^2\text{fm}^2$, which fairly agrees with the observed value  4.07(3) $e^2\text{fm}^2$~\cite{stone_table_2005}.  
As for the excited $3/2^-$ states, the GCM Brink and OCM both predict larger radii than the ground
state.  In particular, the radius of the $3/2_3^-$ state is approximately 3.0 fm and it is comparable with
that of the Hoyle state. The radius of this state from AMD is also relatively large indicating the
pronounced clustering of this state. 

\begin{table}[!htbp]
 \centering
 \caption{\label{tb_monopole} Calculated values of isoscalar monopole transition  strengths in the
 unit of $\rm fm^4$. The
 corresponding results from AMD~\cite{suhara_cluster_2012}, OCM~\cite{yamada_++t_2010}, and experiment~\cite{kawabata_cluster_2007} are also shown for  comparison. } 
 \begin{ruledtabular}
 \begin{center}
  \begin{tabular}{ccccc}
   & GCM Brink  & AMD   & OCM & Experiment   \\	\hline
   $3/2_1^- \to  3/2_2^-$ &  8.42 & 2.5  & $-$  &  \textless\ 9     \\   
   $3/2_1^- \to  3/2_3^-$ &  147  &  150  &  92 &96 $\pm$ 16     \\
  \end{tabular}
 \end{center}
 \end{ruledtabular}
\end{table} 

As already discussed in Refs.~\cite{kawabata_cluster_2007,yamada_monopole_2008}, the enhanced monopole transition strength from the
ground state is a good measure of the clustering of the excited states, and the $3/2^-_3$ state is
known to have fairly large transition strength. 
In Table~\ref{tb_monopole}, we compare our results of the isoscalar monopole transition strengths
with those from AMD, OCM, and experiment. The observed large ISM transition strength~\cite{kawabata_cluster_2007},
$B(ISM; 3/2_1^-\to3/2_{3}^-$)=96 $\pm$ 16 fm$^4$, provides us with a strong support for the
developed cluster structure of 
$3/2_{3}^-$ state. The AMD and OCM confirmed the large ISM transition strength successively. Now,
in our calculations, we also obtained a quite large ISM transition strength, 
$B(ISM; 3/2_1^-\to3/2_{3}^-$)=147 fm$^4$, which is consistent with the experimental data and 
quite close to the AMD result. We also note that the calculated ISM transition strength from the
ground state to the second $ 3/2^-$ state is 8.42 fm$^4$, which 
is also consistent with the experiment (\textless\ 9 fm$^4$).  
From these observations, {\it i.e.} the large radius and ISM transition strength, the AMD, OCM,
and the present GCM Brink reach the same conclusion; the $3/2_3^-$ state is a very developed cluster
state of \bor.

\begin{figure}[h]
 \centering
 \includegraphics[width=0.9\hsize]{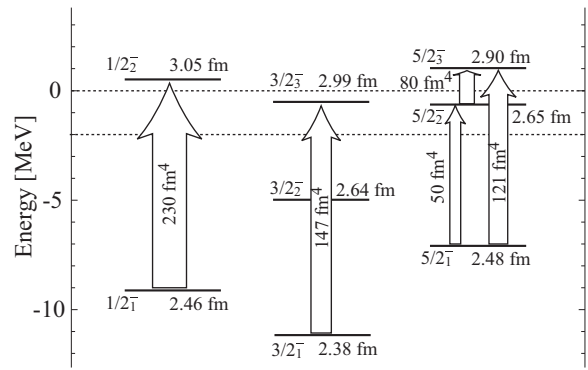}
 \caption{ \label{fig_neg} The GCM-Brink energy levels, r.m.s radii for the mass distributions
 (right side of the energy levels), and the isoscalar monopole transition strengths (stronger than 30 fm$^4$) for
 the negative-parity states in \bor. The above and below dash lines are corresponding to the
 2$\alpha$+$t$  threshold energy and $^{7}$Li($3/2^-$)+$t$  threshold energy, respectively. Units for the energy and radius are MeV and fm, respectively. } 
\end{figure}

Besides the well-confirmed $3/2_3^-$ cluster state, there also exist several $1/2^-$ and $5/2^-$
states around the 2$\alpha+t$ or  $^{7}{\rm Li}+t$  threshold energies, which are also the
candidates of the  developed cluster states. We expect that the ISM transitions from the low-lying
$1/2^-$ and $5/2^-$ states serve as a measure of the clustering. 
Figure~\ref{fig_neg} summarizes the calculated energies, r.m.s radii and ISM transition strengths
of the $1/2^-$ and $5/2^-$ states together with those for the $3/2^-$ states. Firstly, it is noted 
that the radii of the low-lying $1/2^-_1$ and $5/2^-_1$ states are as small as the ground state
indicating their compact shell-model nature as already pointed out by Nishioka {\it et al.}~\cite{nishioka_structure_1979}.
Compared to these compact states, the $1/2^-$ and $5/2^-$ states around the $2\alpha+t$ threshold
have larger radii comparable with or larger than the $3/2^-_3$ state indicating their dilute
clustering. Moreover, it must be noted that the ISM transition strengths to these dilute states
are considerably enhanced. Although it is  difficult to directly measure these quantities, we
consider that they are important signature of the cluster development. Thus, the present result
suggests that $1/2^-_2$ and $5/2^-_3$ states also have dilute cluster structure. 

\begin{table}[htbp]
 \centering
 \caption{\label{tab_neg_be2}  Comparison of the $B(E2)$ strengths of negative-parity states
 between the GCM Brink, AMD~\cite{kawabata_dilute_2007,suhara_cluster_2012}  calculations and the observation~\cite{kelley_energy_2012}. The unit is  $e^2\rm fm^4$.}
 \begin{center}
 \begin{ruledtabular}
  \begin{tabular}{   c  c  c   c  }
   Transition & GCM Brink  & AMD   & Experiment   \\	\hline
 $1/2_1^- \to  3/2_1^-$ & 4.80 & 4.6  & 5.2 $\pm$ 0.8   \\     
 $1/2_1^- \to  3/2_2^-$ &15.5&  13.4 & $116_{-26}^{+233}$   \\     
 $3/2_1^- \to  3/2_2^-$ & 0.07  &   0.02  &   $0.83_{-0.51}^{+0.71}$ \\  
 $3/2_1^- \to  3/2_3^-$ & 1.5  &   0.84 &$-$   \\   
 $3/2_1^- \to  5/2_1^-$ &  14.3  & 13.8  &  13.3 $\pm$ 4.8     \\    
 $3/2_1^- \to  5/2_2^-$ &  2.7  &  0.6 &  $\leq$~0.02     \\   
 $5/2_1^- \to  5/2_2^-$ & 1.6 &$-$ &   0.49 $\pm$ 0.39 \\  
  \end{tabular}
 \end{ruledtabular}
 \end{center}
\end{table} 

We next discuss the $B(E2)$ transition strengths which are useful to test the accuracy of the present
calculation and to identify possible band structure. In Table~\ref{tab_neg_be2}, several $B(E2)$
transition strengths between low-lying states are listed to compare the theoretical results with
the observations. Basically, the present GCM Brink results are qualitatively consistent with the
AMD results. Namely, the transitions $3/2^-_1 \to 5/2^-_1$ and $1/2^-_1 \to 3/2^-_2$  are
enhanced, while others are not.  This characteristics reasonably agrees with the observation
indicating that the present calculation is reliable. 

\begin{figure}[h]
	\centering
	\includegraphics[width=0.9\hsize]{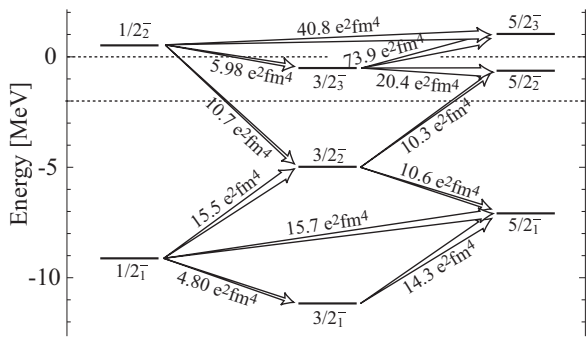}
	\caption{\label{fig_be2_neg} Calculated $B(E2)$ transition strengths between the negative-parity
		states in \bor. The transition strengths lower than 4.80 $e^2\rm fm^4$ are not shown.}
\end{figure}

Then, we discuss the possible band structure from the systematics of the $B(E2)$ strengths
summarized in  Fig.~\ref{fig_be2_neg}. There are several points to be noted in this
figure. Firstly, the transition strength $1/2^-_1\to 3/2^-_2$ is strong, while that for $1/2^-_1
\to 3/2^-_1$ is not. This is because the $1/2^-_1$ and $3/2^-_2$ states are dominated by the
$K^\pi=1/2^-$ component, while  the $3/2^-_2$ state is dominated by the $K^\pi=3/2^-$ component. On the
other hand, the $5/2^-$ states are the admixture of the $K^\pi=3/2^-$ and $1/2^-$ components.  This
explains, for example, why the $5/2^-_1$ state has strong transitions from all of the $1/2^-_1$,
$3/2^-_1$ and $3/2^-_2$ states.  Despite of the $K^\pi=3/2^-$ and $5/2^-$ admixture in the $5/2^-$
states, we suggest a $K^\pi=3/2^-$ band composed of the $3/2^-_1$ and $5/2_1^-$ states, and a $K^\pi=1/2^-$
band composed of the $1/2^-_1$  , $3/2^-_2$  and $5/2^-_2$ states. 

\begin{figure}[h]
	\centering
	\includegraphics[width=0.9\hsize]{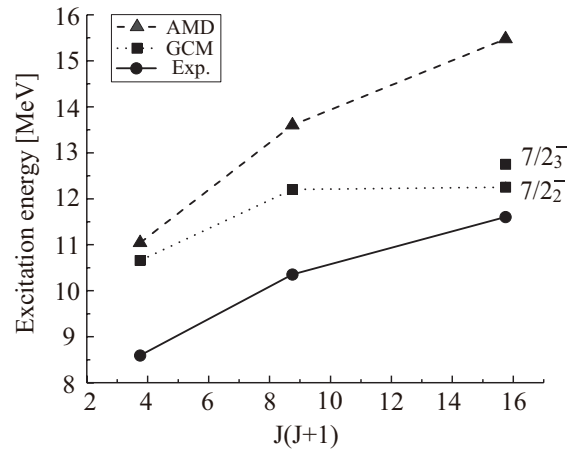}
	\caption{ \label{fig_band} Comparison of the calculated $K^\pi=3/2^-$ bands with GCM Brink ($3/2^-_3$,  $5/2^-_3$, and $7/2^-_2$), AMD ($3/2^-_3$,  $5/2^-_3$, and $7/2^-_3$), and the experimental band ($3/2^-_3$,  $5/2^-_3$, and $7/2^-_3$) suggested in Ref.~\cite{yamaguchi__2011}.}
\end{figure}

Secondly, we see that the $1/2_2^- $, $3/2_3^-$, and $5/2_3^-$ states, which we classified as the 
pronounced cluster states, have large transition strengths because of their large radii. 
It can be seen that the transition $1/2_2^- \to 3/2^-_3$
is rather weak, while $1/2_2^- \to 5/2^-_3$ is strong. Again, it is explained by the $K^\pi=3/2^-$
dominance in the $3/2^-_2$ state, and the $K^\pi=1/2^-,\ 3/2^-$ admixture in the $5/2^-$ states. 

Furthermore, it is found that the transition $3/2_3^- \to 5/2_3^-$ is considerably enhanced. The
AMD calculation also found the significantly large transition between $3/2_3^-$ and $5/2_3^-$
states (and also the $7/2_3^-$ and $9/2_3^-$ states), and hence they suggested a possible band
formation built on the $3/2^-_3$ state as shown in Fig. \ref{fig_band}.  In the present
calculation, we find two possible candidates of the $7/2^-$ states which can be assinged as the
band member state. The $7/2^-_2$ and $7/2^-_3$ are located at 12.25 and 12.75 MeV,
respectively. Their $E2$ transition strengths from the $5/2^-_3$ state are 26.6 and 5.8
$e^2\rm fm^4$, and those from  the $3/2^-_3$ state are 18.3 and 23.4 $e^2\rm fm^4$. However,
because of the coupling with the continuum, present results for $7/2^-$ states are somewhat
ambiguous. Experimentally, by the resonant scattering, a possible candidate for this rotational
band was reported by Yamaguchi {\it et al.} \cite{yamaguchi__2011}.

Finally, we note that the compact shell-model states ($1/2^-_1$, $3/2^-_1$ and $5/2^-_1$ states)
and the dilute cluster states ($1/2^-_2$, $3/2^-_3$, and $5/2^-_3$ states) are disconnected by the
$E2$ transitions due to their considerably different structure. We also note that the
$3/2^-_2$ and $5/2^-_2$ states have strong transition with both of the compact shell-model and
dilute cluster states. This may be because of their transient character between shell and cluster
as seen in their intermediate radius sizes.   

\subsubsection{distribution of $t$ cluster}

\begin{figure*}[!t]
	\centering
	\includegraphics[width=0.9\hsize]{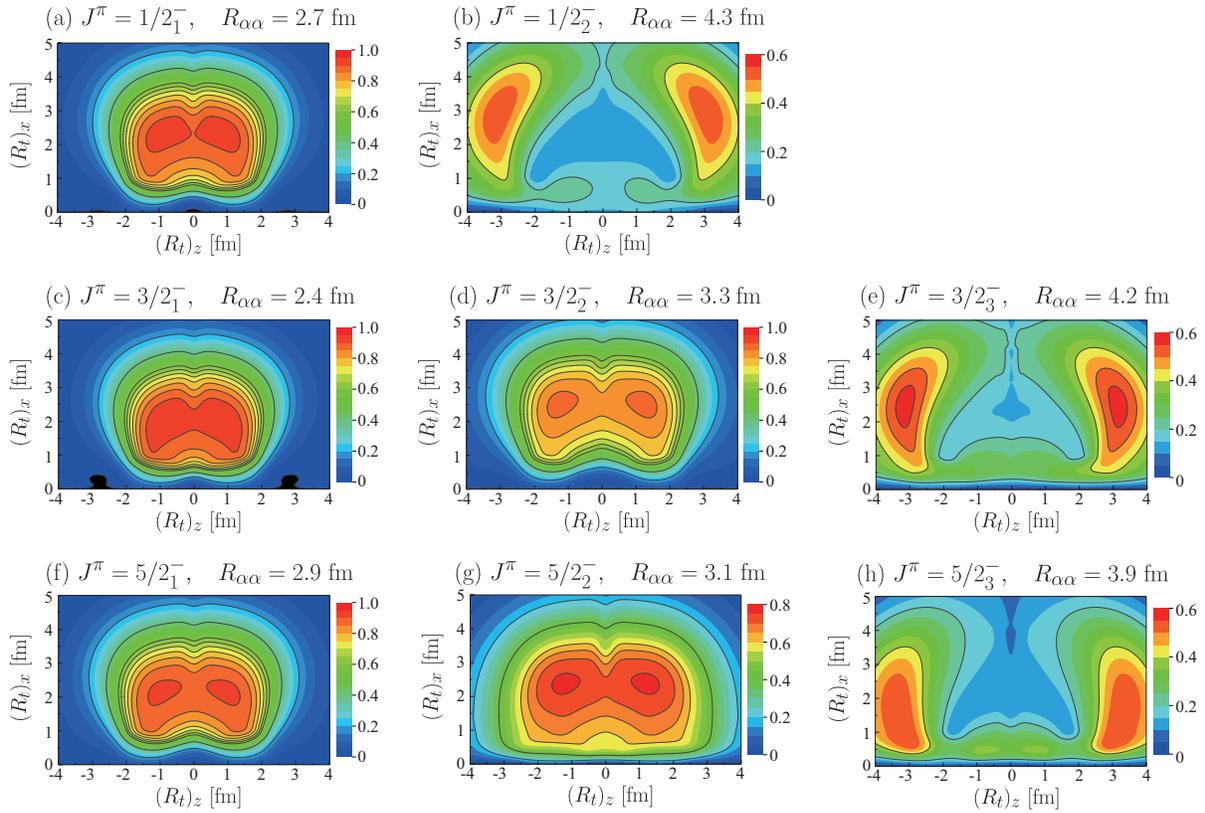}
	\caption{\label{fig_overlap} Contour plots of the squared overlap $O(R_{\alpha\alpha},\bm R_t)$
		defined by Eq. (\ref{overlapr}) for the $ J^{\pi}=1/2^-$, $3/2^-$, and $5/2^-$ states as a function
		of $\bm R_t$. The horizontal axis corresponds to the $z$ component of $\bm R_t$, while the
		vertical axis corresponds to the $x$ component. The optimum distance between 2$\alpha$ 
		clusters which maximizes the squared overlap is also shown.}
\end{figure*}

To study the cluster configuration in the negative-parity states, the squared overlap
${O}(R_{\alpha\alpha},\bm R_t)$ defined by Eq.~(\ref{overlapr}) was calculated.
The distance between 2$\alpha$ clusters $R_{\alpha\alpha}$ is chosen to be an optimum value so 
that the squared overlap is maximized with the proper choice of $\bm R_t$. Thus, by fixing the
$R_{\alpha\alpha}$, the contour plots of $O(R_{\alpha\alpha},\bm R_t)$ are shown in 
Fig.~\ref{fig_overlap}, which show how the $t$ cluster is distributed around the 2$\alpha$
clusters in \bor. 

Figure~\ref{fig_overlap} clearly shows that the optimum values of $R_{\alpha\alpha}$ for the low-lying states
($1/2_1^-$, $3/2_1^-$, and $5/2_1^-$) are smaller than those of the high-lying cluster states.  In
these low-lying states, we also see that the distribution of the $t$ cluster is confined to a
rather small region reflecting the compact shell-model structure.  For example, in case of the
ground state, the squared overlap has the maximum value of  ${O}(R_{\alpha\alpha},\bm R_t)=0.93$ with
$R_{\alpha\alpha}=2.4$ fm and $|\bm R_t|=2.2$ fm.  This means that the ground state can be
reasonably approximated by a single Brink wave function with small inter-cluster distance.

The high-lying cluster states ($1/2_2^-$, $3/2_3^-$, and $5/2_3^-$) show quite contrasting
characteristics. Firstly, they all have relatively large optimum $R_{\alpha\alpha}$, which are
larger than 4 fm. Moreover, the $t$ cluster has a wider distribution. For example, in the case of
the gas-like cluster state $3/2_3^-$, the squared overlap has the maximum value of 
$O(R_{\alpha\alpha},\bm R_t)=0.57$ with $R_{\alpha\alpha}=4.2$ fm and $|\bm R_t|=4.0$ fm. The
smaller value of the maximum overlap and larger value of the inter-cluster distance indicate that
the state cannot be approximated by a single cluster configuration due to its dilute gas-like
nature. It must be noted that the $1/2^-_2$ and $5/2_3^-$ states also show quite similar
distributions, and hence, we conclude that they also have dilute cluster structure. 

As a whole for the negative-parity states, with the increase of excitation energy, we see that
the compact shell-model states evolve to the developed cluster states. The $3/2_2^-$ and $5/2_2^-$
can be regarded as the intermediate states, which have larger $R_{\alpha\alpha}$  and wider
$t$-cluster distributions than the ground state.

\subsection{Positive-parity states in \bor\ }
\subsubsection{radii and transition strengths}
\begin{figure}[!t]
	\centering
	\includegraphics[width=0.9\hsize]{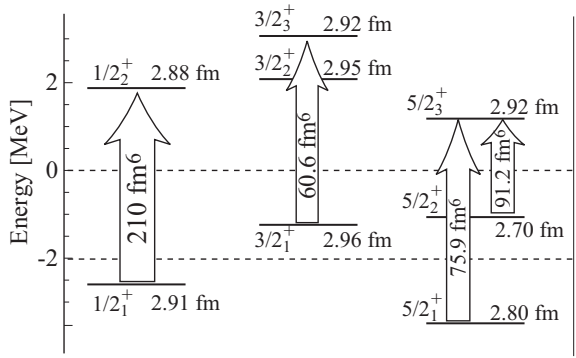}
	\caption{ \label{fig_pos} The GCM-Brink energy levels, r.m.s radii for the mass distributions
		(right side of the energy levels), and the isoscalar monopole transition strengths (stronger than 60 fm$^4$) for
		the positive-parity states in \bor. The above and below dash lines are corresponding to the
		2$\alpha$+$t$  threshold energy and $^{7}$Li($3/2^-$)+$t$  threshold energy, respectively. The
		Units for the energy and radius are MeV and fm, respectively.} 
\end{figure}

In general, compared with the negative-parity states, the calculated positive-parity states are
located at relatively high energy region, and are all around the 2$\alpha+t$ or 
$^{7}{\rm Li}(3/2^-)+t$  threshold energies. Therefore, it seems that more positive-parity states
are promising to have developed cluster structure. 

\begin{table}[h]
	\centering
	\caption{\label{tb_posrad} The r.m.s radii for mass distributions of $1/2^+$  states from  GCM
		Brink and OCM calculations~\cite{yamada_three-body_2012}. The unit is fm.}  
	\begin{center}
		\begin{ruledtabular}
			\begin{tabular}{ccc}
				State & GCM Brink& OCM     \\	\hline
				$1/2_1^+ $ & 2.91 &  2.82     \\  
				$1/2_2^+$ &  2.88 & 5.93        \\ 
			\end{tabular}
		\end{ruledtabular}
	\end{center}
\end{table}

Similar to the nagative-parity case, we expect that the well-developed positive-parity cluster
states should also have pronounced monopole transition strengths from the $1/2^+_1$, $3/2^+_1$, and
$5/2^+_1$ states. Figure~\ref{fig_pos} shows the calculated energy levels, r.m.s radii, and the
ISM transition strengths for the positive-parity states. For the $J^\pi=1/2^+$ states, a 
very strong ISM transition strength between $1/2^+_1$ and $1/2^+_2$ states was obtained, which
reaches 210 fm$^4$ and even much stronger than the ISM strength between the ground and $3/2^-_3$
states. The strong ISM transitions also occur between the $3/2^+_1$ state and $3/2^+_3$ state,
and between the $5/2^+_1$ state and $5/2^+_3$ state. Indeed, from the view of radius, except for
the $5/2^+_2$ state, all the other states have very large radii comparable with that of the
developed $3/2^-_3$ cluster state, which indicate that these most positive-parity states can be
considered as the candidates of cluster states in \bor. 

In the OCM calculation, the obtained $1/2^+_2$ state is a strong candidate for the Hoyle-analogue
state in which all clusters occupy the dilute $s$-wave state. An important feature for this state 
obtained by the OCM study is its extremely large radius, 5.93 fm. However, in the present
calculation, the radius of the $1/2^+_2$ state is not so large and even smaller than that of the
$1/2^+_1$ state, which can be seen in Table~\ref{tb_posrad}. This means that the  $1/2^+_2$ states
obtained by the GCM Brink and OCM are incompatible, which will be discussed later.

\begin{figure}[h]
	\centering
	\includegraphics[width=0.9\hsize]{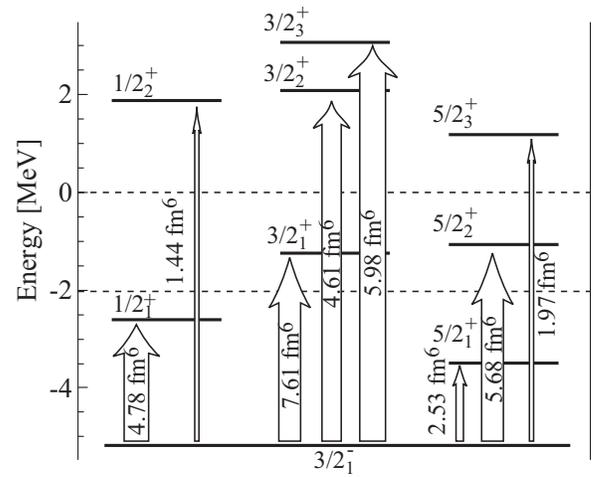}
	\caption{ \label{fig_posdip}  ISD transitions from the ground state to the positive-parity states in \bor. }  
\end{figure}

Besides the ISM transition, the ISD transition is another useful probe for the clustering.
Since most of the calculated  positive-parity states have large radii, we expect that the
isoscalar dipole transition can be enhanced. Unexpectedly, the obtained ISD transition strengths
are not enhanced as shown in Fig.~\ref{fig_posdip}. 
The calculated transition strengths range from 1.4 to 7.6 fm$^6$, which are not as strong as
the single-particle estimate 21.6 $\rm fm^6$. 
These strengths should be directly compared with the
experimental data to test the wave functions of the present calculation. There must be some kind
of underlying theoretical reason for these relatively weak but distinct strengths, which will be explored in the
future.
\begin{figure}[!b]
	\centering
	\includegraphics[width=0.9\hsize]{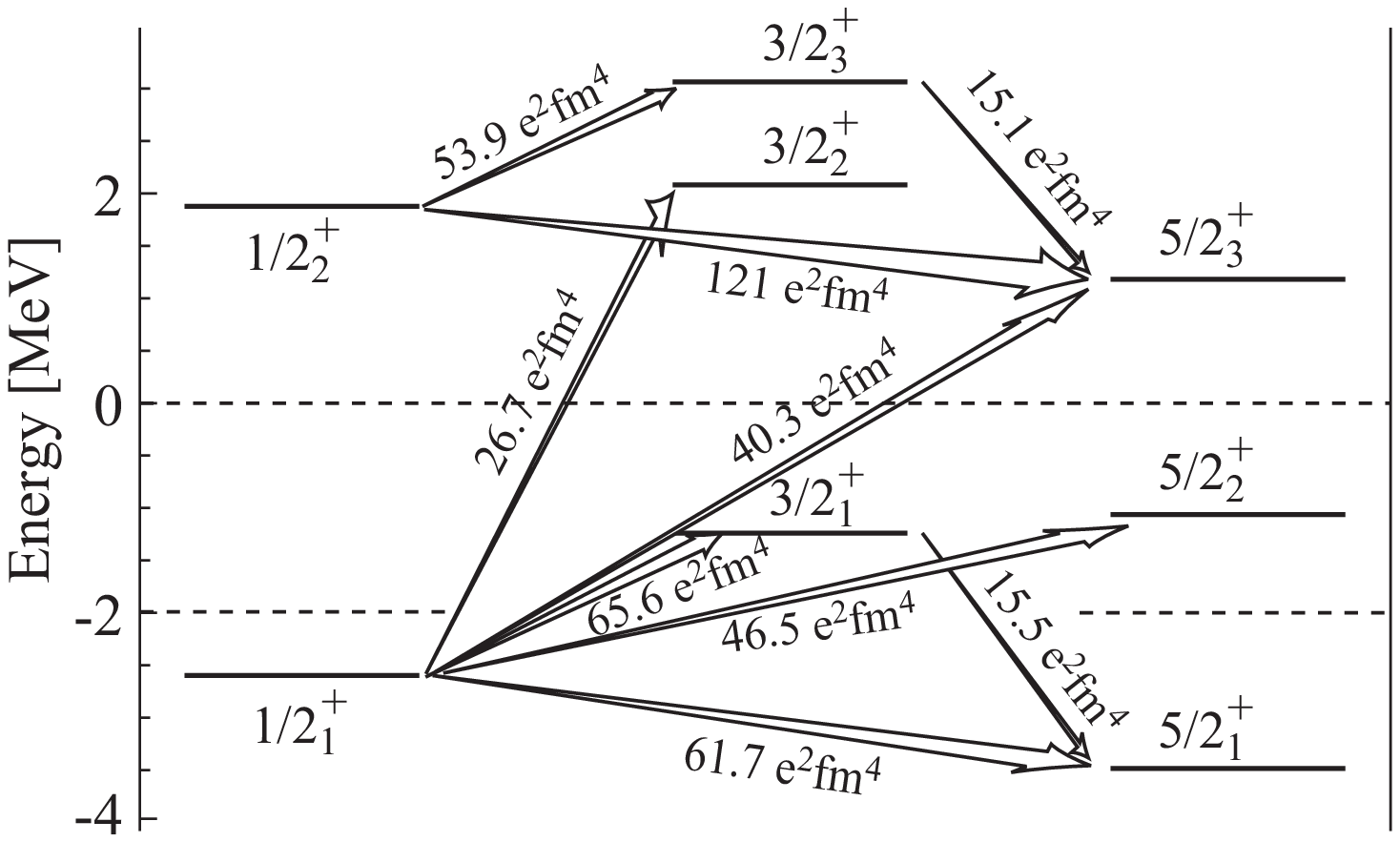}
	\caption{ \label{fig_be2_pos} Positive-parity energy levels with the $B(E2)$ transition strengths
		(stronger than 15 $e^2$fm$^4$).} 
\end{figure}

Next, we focus on the  $B(E2)$ transition strengths shown in Fig.~\ref{fig_be2_pos}. Clearly, the
excited states are classified into two groups; ($1/2^+_1$, $3/2^+_1$, $5/2^+_1$) and  ($1/2^+_2$,
$3/2^+_3$, $5/2^+_3$), which are mutually connected with strong transitions. 
These two groups of states show quite similar pattern. For example, the strengths
$B(E2; 1/2^+_1\to 3/2^+_1)$ and $B(E2; 1/2^+_2\to3/2^+_3)$ are both quite strong and larger than
50 $e^2\rm fm^4$ while the strengths $B(E2; 3/2^+_1\to5/2^+_1)$ and $B(E2; 3/2^+_3\to5/2^+_3)$
are about 15 $e^2\rm fm^4$. The $3/2^+_2$ state is admixture of the $K^\pi=1/2^+$ and $3/2^+$
components, which explains the non-negligible transition strength between $1/2^+_1$ and $3/2^+_2$
states.  Another fact to be reminded is, as we already mentioned, that there are also very strong
monopole transitions between the corresponding states.

\subsubsection{distribution of $t$ cluster}
\begin{figure*}[!t]
	\centering
	\includegraphics[width=0.9\hsize]{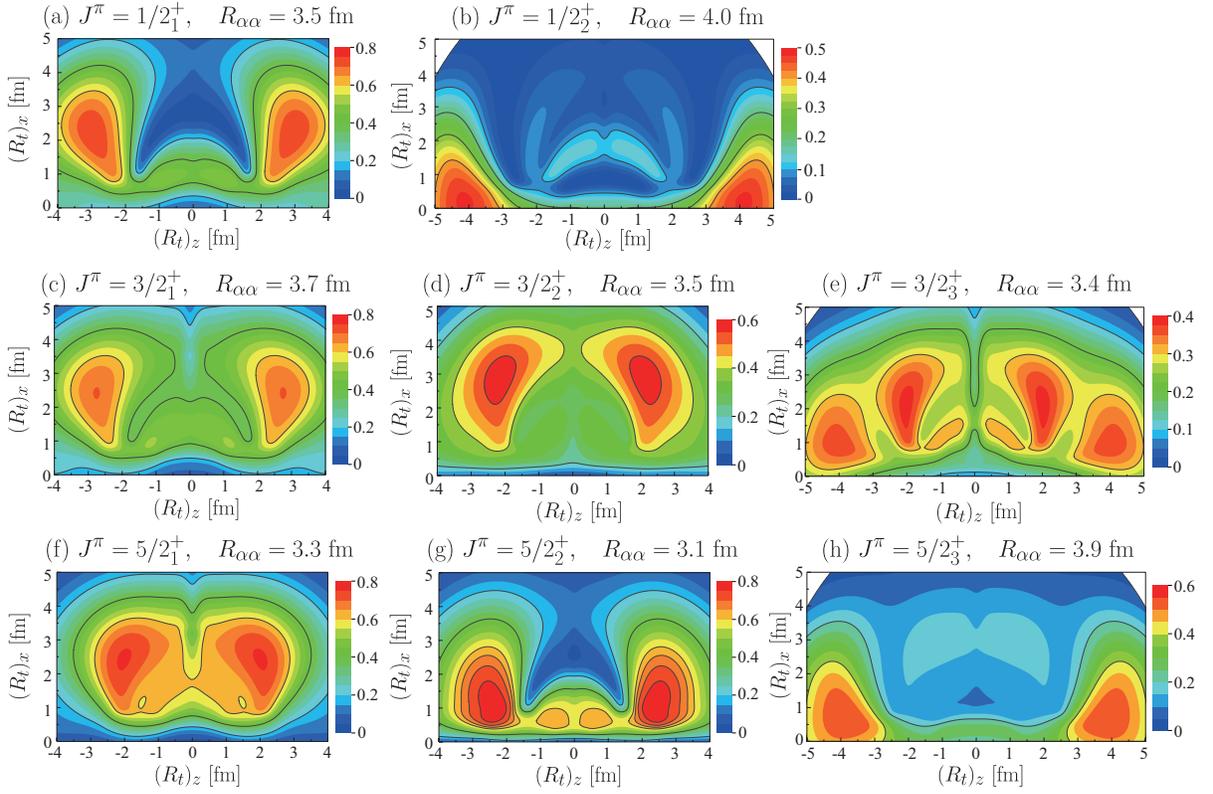}
	\caption{\label{fig_overlap_pos} Contour plots of the squared overlap $O(R_{\alpha\alpha},\bm
		R_t)$  defined by Eq.~(\ref{overlapr}) for the $ J^{\pi}=1/2^+$, $3/2^+$, and $5/2^+$ states as
		a function  of $\bm R_t$. The horizontal axis corresponds to the $z$ component of $\bm R_t$, while
		the vertical axis corresponds to the $x$ component. The optimum distance between 2$\alpha$  
		clusters which maximizes the squared overlap is also shown.}
\end{figure*}

Figure~\ref{fig_overlap_pos} is the contour plot of the squared overlap between the Brink wave
function and GCM wave function for the positive-parity states. Overall,  most of the
positive-parity states show the pronounced clustering as expected, in which the 
optimum  2$\alpha$ distance $R_{\alpha\alpha}$ is very large and the triton cluster has relatively
broad distribution. We see that there are several different patterns in the $t$ cluster
distributions.

First and most obviously, the obtained $1/2^+_2$ state has a linear-chain structure. In
Fig~\ref{fig_overlap_pos} (b), it can be seen that the squared overlap is largest at each end of the $z$ 
axis ($(R_t)_x =0$ and $|(R_t)_z| = 4.3$ fm), but rapidly decreases in other region. This indicates
that the $t$ cluster is mostly confined in the linearly aligned $\alpha$-$\alpha$-$t$ or
$t$-$\alpha$-$\alpha$ configuration. The squared overlap has a maximum value 0.476 when
$R_{\alpha\alpha}=4.0$ fm and $|\bm{R}_t|=4.3$ fm, which approximately corresponds to touching three clusters. This
explains why the r.m.s radius of the $1/2^+_2$ state is not so large and even smaller than that of the $1/2^+_1$ state. It is noted that the $1/2^+_1$ state obtained by the AMD calculation \cite{suhara_cluster_2012} has quite similar
linear-chain structure. The reported excitation energy 13.6 MeV is also very close to the present
result 13.0 MeV. Therefore, we may be able to conclude that our $1/2^+_2$ state and the $1/2^+_1$
state reported by AMD calculation are identical state. The reason for the missing $1/2^+$ state in
the AMD calculation corresponding to our $1/2^+_1$ state may be due to the calculation procedure
applied in the AMD study. They performed the energy variation before the angular momentum
projection, which occasionally fails to describe the energetically unfavored states.

In contrast to the AMD result, the  OCM calculation reported no linear-chain configurations, but
reported the dilute gas-like charactor of the $1/2^+_2$ state. 
Therefore, it is a quite interesting and important problem to clarify the character of the
$1/2^+_2$ state. 

Moreover, in addition to the $1/2^+_2$ state, the $5/2^+_3$ state also shows a kind of bent linear-chain structure, e.g., the maximum square overlap 0.54 appears at $(R_t)_x = 0.8$ fm and $|(R_t)_z|= 3.9$ fm. In fact, we have mentioned that these two states have been strongly connected by $E2$ transition and the strength $B(E2; 1/2^+_2\to5/2^+_3)$ is as high as 121 $e^2$fm$^4$. In Fig.~\ref{fig_overlap_pos}, from the distributions of triton cluster, it can be seen that the isosceles triangle shape is not favored for the cluster states in \bor. In addition, the $5/2^+_1$ state is shown having a compact cluster structure, considering it is the lowest positive-parity state in our calculations and it has a relatively small radius, which can be considered as a transient state in \bor. 

\section{\label{sec4}Summary}
In this study, we investigated cluster states in \bor\ using the GCM Brink wave
functions. Firstly, with a proper choice of the parameters of the effective interaction, the energy spectrum including the negative-parity and positive-parity states were reproduced in the present framework. Moreover, based on the calculated radii and monopole transition strengths, several cluster states around the 2$\alpha$+$t$  threshold energy were suggested in \bor. Some rotational bands were also discussed according to the obtained $B(E2)$ transition strengths. Furthermore,  by calculating the squared overlap between GCM Brink and projected Brink wave functions,  the $t$-cluster distributions were shown for different states and the proposed cluster states were further supported.   
 
For negative-parity states in \bor, the $1/2_2^-$ and $5/2_3^-$ states as well as the well-known $3/2_3^-$
state were shown to be the well-developed cluster states. The obtained $3/2_3^-$ and $5/2_3^-$ states were strongly connected by $E2$ transitions and a rotational cluster band $K^\pi$=$3/2^-$ was proposed. In the description of the negative-parity states of \bor, 
GCM Brink, AMD, and OCM actually are quite consistent and they provide us with similar results from the comparisons.
As for the positive-parity states, most states around 2$\alpha$+$t$  threshold show some features of cluster structure.  Most importantly, it is found that the obtained $1/2_2^+$ state can be considered as a linear-chain-like state, which is consistent
with the $1/2_1^+$ state from AMD but very different from the obtained $1/2_2^+$ state from
OCM. To clarify further the characters of positive-parity cluster states especially the $1/2_2^+$ cluster state, the
experimental data for the isoscalar dipole transition strengths were highly required.  

\begin{acknowledgments}
The authors acknowledge fruitful discussions with Prof. T. Kawabata, Prof. T. Yamada, Prof. Y. Funaki, and Dr. T. Suhara. 
This work was supported by JSPS KAKENHI Grant Numbers 17K1426207 (Grant-in-Aid for Young Scientists (B)) and   
16K05339 (Grant-in-Aid for Scientific Research (C)). Numerical computation in this work was carried out at the Yukawa Institute Computer Facility in Kyoto University.
\end{acknowledgments}


\begin{thebibliography}{34}%
	\makeatletter
	\providecommand \@ifxundefined [1]{%
		\@ifx{#1\undefined}
	}%
	\providecommand \@ifnum [1]{%
		\ifnum #1\expandafter \@firstoftwo
		\else \expandafter \@secondoftwo
		\fi
	}%
	\providecommand \@ifx [1]{%
		\ifx #1\expandafter \@firstoftwo
		\else \expandafter \@secondoftwo
		\fi
	}%
	\providecommand \natexlab [1]{#1}%
	\providecommand \enquote  [1]{``#1''}%
	\providecommand \bibnamefont  [1]{#1}%
	\providecommand \bibfnamefont [1]{#1}%
	\providecommand \citenamefont [1]{#1}%
	\providecommand \href@noop [0]{\@secondoftwo}%
	\providecommand \href [0]{\begingroup \@sanitize@url \@href}%
	\providecommand \@href[1]{\@@startlink{#1}\@@href}%
	\providecommand \@@href[1]{\endgroup#1\@@endlink}%
	\providecommand \@sanitize@url [0]{\catcode `\\12\catcode `\$12\catcode
		`\&12\catcode `\#12\catcode `\^12\catcode `\_12\catcode `\%12\relax}%
	\providecommand \@@startlink[1]{}%
	\providecommand \@@endlink[0]{}%
	\providecommand \url  [0]{\begingroup\@sanitize@url \@url }%
	\providecommand \@url [1]{\endgroup\@href {#1}{\urlprefix }}%
	\providecommand \urlprefix  [0]{URL }%
	\providecommand \Eprint [0]{\href }%
	\providecommand \doibase [0]{http://dx.doi.org/}%
	\providecommand \selectlanguage [0]{\@gobble}%
	\providecommand \bibinfo  [0]{\@secondoftwo}%
	\providecommand \bibfield  [0]{\@secondoftwo}%
	\providecommand \translation [1]{[#1]}%
	\providecommand \BibitemOpen [0]{}%
	\providecommand \bibitemStop [0]{}%
	\providecommand \bibitemNoStop [0]{.\EOS\space}%
	\providecommand \EOS [0]{\spacefactor3000\relax}%
	\providecommand \BibitemShut  [1]{\csname bibitem#1\endcsname}%
	\let\auto@bib@innerbib\@empty
	\bibitem [{\citenamefont {Wildermuth}\ and\ \citenamefont
		{Tang}(1977)}]{wildermuth_unified_1977}%
	\BibitemOpen
	\bibfield  {author} {\bibinfo {author} {\bibfnamefont {K.}~\bibnamefont
			{Wildermuth}}\ and\ \bibinfo {author} {\bibfnamefont {Y.~C.}\ \bibnamefont
			{Tang}},\ }\href@noop {} {\emph {\bibinfo {title} {A unified theory of the
				nucleus}}}\ (\bibinfo  {publisher} {Vieweg},\ \bibinfo {year}
	{1977})\BibitemShut {NoStop}%
	\bibitem [{\citenamefont {Fujiwara}\ \emph {et~al.}(1980)\citenamefont
		{Fujiwara}, \citenamefont {Horiuchi}, \citenamefont {Ikeda}, \citenamefont
		{Kamimura} \emph {et~al.}}]{fujiwara_chapter_1980}%
	\BibitemOpen
	\bibfield  {author} {\bibinfo {author} {\bibfnamefont {Y.}~\bibnamefont
			{Fujiwara}}, \bibinfo {author} {\bibfnamefont {H.}~\bibnamefont {Horiuchi}},
		\bibinfo {author} {\bibfnamefont {K.}~\bibnamefont {Ikeda}}, \bibinfo
		{author} {\bibfnamefont {M.}~\bibnamefont {Kamimura}},  \emph {et~al.},\
	}\href@noop {} {\bibfield  {journal} {\bibinfo  {journal} {Prog. Theor. Phys.
				Suppl.}\ }\textbf {\bibinfo {volume} {68}},\ \bibinfo {pages} {29} (\bibinfo
		{year} {1980})}\BibitemShut {NoStop}%
	\bibitem [{\citenamefont {Kanada-En'yo}\ \emph {et~al.}(2003)\citenamefont
		{Kanada-En'yo}, \citenamefont {Kimura},\ and\ \citenamefont
		{Horiuchi}}]{kanada-enyo_antisymmetrized_2003}%
	\BibitemOpen
	\bibfield  {author} {\bibinfo {author} {\bibfnamefont {Y.}~\bibnamefont
			{Kanada-En'yo}}, \bibinfo {author} {\bibfnamefont {M.}~\bibnamefont
			{Kimura}}, \ and\ \bibinfo {author} {\bibfnamefont {H.}~\bibnamefont
			{Horiuchi}},\ }\href@noop {} {\bibfield  {journal} {\bibinfo  {journal} {C.
				R. Phys.}\ }\textbf {\bibinfo {volume} {4}},\ \bibinfo {pages} {497}
		(\bibinfo {year} {2003})}\BibitemShut {NoStop}%
	\bibitem [{\citenamefont {Freer}(2007)}]{freer_clustered_2007}%
	\BibitemOpen
	\bibfield  {author} {\bibinfo {author} {\bibfnamefont {M.}~\bibnamefont
			{Freer}},\ }\href@noop {} {\bibfield  {journal} {\bibinfo  {journal} {Rep.
				Prog. Phys.}\ }\textbf {\bibinfo {volume} {70}},\ \bibinfo {pages} {2149}
		(\bibinfo {year} {2007})}\BibitemShut {NoStop}%
	\bibitem [{\citenamefont {Horiuchi}\ \emph {et~al.}(2012)\citenamefont
		{Horiuchi}, \citenamefont {Ikeda},\ and\ \citenamefont
		{Kat\={o}}}]{horiuchi_recent_2012}%
	\BibitemOpen
	\bibfield  {author} {\bibinfo {author} {\bibfnamefont {H.}~\bibnamefont
			{Horiuchi}}, \bibinfo {author} {\bibfnamefont {K.}~\bibnamefont {Ikeda}}, \
		and\ \bibinfo {author} {\bibfnamefont {K.}~\bibnamefont {Kat\={o}}},\
	}\href@noop {} {\bibfield  {journal} {\bibinfo  {journal} {Prog. Theor. Phys.
				Suppl.}\ }\textbf {\bibinfo {volume} {192}},\ \bibinfo {pages} {1} (\bibinfo
		{year} {2012})}\BibitemShut {NoStop}%
	\bibitem [{\citenamefont {Kimura}\ \emph {et~al.}(2016)\citenamefont {Kimura},
		\citenamefont {Suhara},\ and\ \citenamefont
		{Kanada-En'yo}}]{kimura_antisymmetrized_2016}%
	\BibitemOpen
	\bibfield  {author} {\bibinfo {author} {\bibfnamefont {M.}~\bibnamefont
			{Kimura}}, \bibinfo {author} {\bibfnamefont {T.}~\bibnamefont {Suhara}}, \
		and\ \bibinfo {author} {\bibfnamefont {Y.}~\bibnamefont {Kanada-En'yo}},\
	}\href@noop {} {\bibfield  {journal} {\bibinfo  {journal} {Eur. Phys. J. A}\
		}\textbf {\bibinfo {volume} {52}},\ \bibinfo {pages} {373} (\bibinfo {year}
		{2016})}\BibitemShut {NoStop}%
	\bibitem [{\citenamefont {Tohsaki}\ \emph {et~al.}(2001)\citenamefont
		{Tohsaki}, \citenamefont {Horiuchi}, \citenamefont {Schuck},\ and\
		\citenamefont {R\"{o}pke}}]{tohsaki_alpha_2001}%
	\BibitemOpen
	\bibfield  {author} {\bibinfo {author} {\bibfnamefont {A.}~\bibnamefont
			{Tohsaki}}, \bibinfo {author} {\bibfnamefont {H.}~\bibnamefont {Horiuchi}},
		\bibinfo {author} {\bibfnamefont {P.}~\bibnamefont {Schuck}}, \ and\ \bibinfo
		{author} {\bibfnamefont {G.}~\bibnamefont {R\"{o}pke}},\ }\href@noop {}
	{\bibfield  {journal} {\bibinfo  {journal} {Phys. Rev. Lett.}\ }\textbf
		{\bibinfo {volume} {87}},\ \bibinfo {pages} {192501} (\bibinfo {year}
		{2001})}\BibitemShut {NoStop}%
	\bibitem [{\citenamefont {Tohsaki}\ \emph {et~al.}(2017)\citenamefont
		{Tohsaki}, \citenamefont {Horiuchi}, \citenamefont {Schuck},\ and\
		\citenamefont {R\"{o}pke}}]{tohsaki_colloquium_2017}%
	\BibitemOpen
	\bibfield  {author} {\bibinfo {author} {\bibfnamefont {A.}~\bibnamefont
			{Tohsaki}}, \bibinfo {author} {\bibfnamefont {H.}~\bibnamefont {Horiuchi}},
		\bibinfo {author} {\bibfnamefont {P.}~\bibnamefont {Schuck}}, \ and\ \bibinfo
		{author} {\bibfnamefont {G.}~\bibnamefont {R\"{o}pke}},\ }\href@noop {}
	{\bibfield  {journal} {\bibinfo  {journal} {Rev. Mod. Phys.}\ }\textbf
		{\bibinfo {volume} {89}},\ \bibinfo {pages} {011002} (\bibinfo {year}
		{2017})}\BibitemShut {NoStop}%
	\bibitem [{\citenamefont {von Oertzen}\ \emph {et~al.}(2006)\citenamefont {von
			Oertzen}, \citenamefont {Freer},\ and\ \citenamefont
		{Kanada-En’yo}}]{von_oertzen_nuclear_2006}%
	\BibitemOpen
	\bibfield  {author} {\bibinfo {author} {\bibfnamefont {W.}~\bibnamefont {von
				Oertzen}}, \bibinfo {author} {\bibfnamefont {M.}~\bibnamefont {Freer}}, \
		and\ \bibinfo {author} {\bibfnamefont {Y.}~\bibnamefont {Kanada-En’yo}},\
	}\href@noop {} {\bibfield  {journal} {\bibinfo  {journal} {Phys. Rep.}\
		}\textbf {\bibinfo {volume} {432}},\ \bibinfo {pages} {43} (\bibinfo {year}
		{2006})}\BibitemShut {NoStop}%
	\bibitem [{\citenamefont {Itoh}\ \emph {et~al.}(2011)\citenamefont {Itoh},
		\citenamefont {Akimune}, \citenamefont {Fujiwara}, \citenamefont {Garg} \emph
		{et~al.}}]{itoh_candidate_2011}%
	\BibitemOpen
	\bibfield  {author} {\bibinfo {author} {\bibfnamefont {M.}~\bibnamefont
			{Itoh}}, \bibinfo {author} {\bibfnamefont {H.}~\bibnamefont {Akimune}},
		\bibinfo {author} {\bibfnamefont {M.}~\bibnamefont {Fujiwara}}, \bibinfo
		{author} {\bibfnamefont {U.}~\bibnamefont {Garg}},  \emph {et~al.},\
	}\href@noop {} {\bibfield  {journal} {\bibinfo  {journal} {Phys. Rev. C}\
		}\textbf {\bibinfo {volume} {84}},\ \bibinfo {pages} {054308} (\bibinfo
		{year} {2011})}\BibitemShut {NoStop}%
	\bibitem [{\citenamefont {Yang}\ \emph {et~al.}(2014)\citenamefont {Yang},
		\citenamefont {Ye}, \citenamefont {Li}, \citenamefont {Lou} \emph
		{et~al.}}]{yang_observation_2014}%
	\BibitemOpen
	\bibfield  {author} {\bibinfo {author} {\bibfnamefont {Z.}~\bibnamefont
			{Yang}}, \bibinfo {author} {\bibfnamefont {Y.}~\bibnamefont {Ye}}, \bibinfo
		{author} {\bibfnamefont {Z.}~\bibnamefont {Li}}, \bibinfo {author}
		{\bibfnamefont {J.}~\bibnamefont {Lou}},  \emph {et~al.},\ }\href@noop {}
	{\bibfield  {journal} {\bibinfo  {journal} {Phys. Rev. Lett.}\ }\textbf
		{\bibinfo {volume} {112}},\ \bibinfo {pages} {162501} (\bibinfo {year}
		{2014})}\BibitemShut {NoStop}%
	\bibitem [{\citenamefont {He}\ \emph {et~al.}(2014)\citenamefont {He},
		\citenamefont {Ma}, \citenamefont {Cao}, \citenamefont {Cai},\ and\
		\citenamefont {Zhang}}]{he_giant_2014}%
	\BibitemOpen
	\bibfield  {author} {\bibinfo {author} {\bibfnamefont {W.}~\bibnamefont
			{He}}, \bibinfo {author} {\bibfnamefont {Y.}~\bibnamefont {Ma}}, \bibinfo
		{author} {\bibfnamefont {X.}~\bibnamefont {Cao}}, \bibinfo {author}
		{\bibfnamefont {X.}~\bibnamefont {Cai}}, \ and\ \bibinfo {author}
		{\bibfnamefont {G.}~\bibnamefont {Zhang}},\ }\href@noop {} {\bibfield
		{journal} {\bibinfo  {journal} {Phys. Rev. Lett.}\ }\textbf {\bibinfo
			{volume} {113}},\ \bibinfo {pages} {032506} (\bibinfo {year}
		{2014})}\BibitemShut {NoStop}%
	\bibitem [{\citenamefont {Zhou}\ \emph {et~al.}(2016)\citenamefont {Zhou},
		\citenamefont {Tohsaki}, \citenamefont {Horiuchi},\ and\ \citenamefont
		{Ren}}]{zhou_breathing-like_2016}%
	\BibitemOpen
	\bibfield  {author} {\bibinfo {author} {\bibfnamefont {B.}~\bibnamefont
			{Zhou}}, \bibinfo {author} {\bibfnamefont {A.}~\bibnamefont {Tohsaki}},
		\bibinfo {author} {\bibfnamefont {H.}~\bibnamefont {Horiuchi}}, \ and\
		\bibinfo {author} {\bibfnamefont {Z.}~\bibnamefont {Ren}},\ }\href@noop {}
	{\bibfield  {journal} {\bibinfo  {journal} {Phys. Rev. C}\ }\textbf {\bibinfo
			{volume} {94}},\ \bibinfo {pages} {044319} (\bibinfo {year}
		{2016})}\BibitemShut {NoStop}%
	\bibitem [{\citenamefont {Baba}\ and\ \citenamefont
		{Kimura}(2016)}]{baba_structure_2016}%
	\BibitemOpen
	\bibfield  {author} {\bibinfo {author} {\bibfnamefont {T.}~\bibnamefont
			{Baba}}\ and\ \bibinfo {author} {\bibfnamefont {M.}~\bibnamefont {Kimura}},\
	}\href@noop {} {\bibfield  {journal} {\bibinfo  {journal} {Phys. Rev. C}\
		}\textbf {\bibinfo {volume} {94}},\ \bibinfo {pages} {044303} (\bibinfo
		{year} {2016})}\BibitemShut {NoStop}%
	\bibitem [{\citenamefont {Nishioka}\ \emph {et~al.}(1979)\citenamefont
		{Nishioka}, \citenamefont {Saito},\ and\ \citenamefont
		{Yasuno}}]{nishioka_structure_1979}%
	\BibitemOpen
	\bibfield  {author} {\bibinfo {author} {\bibfnamefont {H.}~\bibnamefont
			{Nishioka}}, \bibinfo {author} {\bibfnamefont {S.}~\bibnamefont {Saito}}, \
		and\ \bibinfo {author} {\bibfnamefont {M.}~\bibnamefont {Yasuno}},\
	}\href@noop {} {\bibfield  {journal} {\bibinfo  {journal} {Prog. Theor.
				Phys.}\ }\textbf {\bibinfo {volume} {62}},\ \bibinfo {pages} {424} (\bibinfo
		{year} {1979})}\BibitemShut {NoStop}%
	\bibitem [{\citenamefont {Suhara}\ and\ \citenamefont
		{Kanada-En'yo}(2012)}]{suhara_cluster_2012}%
	\BibitemOpen
	\bibfield  {author} {\bibinfo {author} {\bibfnamefont {T.}~\bibnamefont
			{Suhara}}\ and\ \bibinfo {author} {\bibfnamefont {Y.}~\bibnamefont
			{Kanada-En'yo}},\ }\href@noop {} {\bibfield  {journal} {\bibinfo  {journal}
			{Phys. Rev. C}\ }\textbf {\bibinfo {volume} {85}},\ \bibinfo {pages} {054320}
		(\bibinfo {year} {2012})}\BibitemShut {NoStop}%
	\bibitem [{\citenamefont {Kawabata}\ \emph
		{et~al.}(2007{\natexlab{a}})\citenamefont {Kawabata}, \citenamefont
		{Akimune}, \citenamefont {Fujita}, \citenamefont {Fujita} \emph
		{et~al.}}]{kawabata_cluster_2007}%
	\BibitemOpen
	\bibfield  {author} {\bibinfo {author} {\bibfnamefont {T.}~\bibnamefont
			{Kawabata}}, \bibinfo {author} {\bibfnamefont {H.}~\bibnamefont {Akimune}},
		\bibinfo {author} {\bibfnamefont {H.}~\bibnamefont {Fujita}}, \bibinfo
		{author} {\bibfnamefont {Y.}~\bibnamefont {Fujita}},  \emph {et~al.},\
	}\href@noop {} {\bibfield  {journal} {\bibinfo  {journal} {Phys. Lett. B}\
		}\textbf {\bibinfo {volume} {646}},\ \bibinfo {pages} {6} (\bibinfo {year}
		{2007}{\natexlab{a}})}\BibitemShut {NoStop}%
	\bibitem [{\citenamefont {Kanada-En'yo}\ and\ \citenamefont
		{Suhara}(2015)}]{kanada-enyo_2alpha+triton_2015}%
	\BibitemOpen
	\bibfield  {author} {\bibinfo {author} {\bibfnamefont {Y.}~\bibnamefont
			{Kanada-En'yo}}\ and\ \bibinfo {author} {\bibfnamefont {T.}~\bibnamefont
			{Suhara}},\ }\href@noop {} {\bibfield  {journal} {\bibinfo  {journal} {Phys.
				Rev. C}\ }\textbf {\bibinfo {volume} {91}},\ \bibinfo {pages} {014316}
		(\bibinfo {year} {2015})}\BibitemShut {NoStop}%
	\bibitem [{\citenamefont {Yamada}\ and\ \citenamefont
		{Funaki}(2010)}]{yamada_++t_2010}%
	\BibitemOpen
	\bibfield  {author} {\bibinfo {author} {\bibfnamefont {T.}~\bibnamefont
			{Yamada}}\ and\ \bibinfo {author} {\bibfnamefont {Y.}~\bibnamefont
			{Funaki}},\ }\href@noop {} {\bibfield  {journal} {\bibinfo  {journal} {Phys.
				Rev. C}\ }\textbf {\bibinfo {volume} {82}},\ \bibinfo {pages} {064315}
		(\bibinfo {year} {2010})}\BibitemShut {NoStop}%
	\bibitem [{\citenamefont {Yamaguchi}\ \emph {et~al.}(2011)\citenamefont
		{Yamaguchi}, \citenamefont {Hashimoto}, \citenamefont {Hayakawa},
		\citenamefont {Binh} \emph {et~al.}}]{yamaguchi__2011}%
	\BibitemOpen
	\bibfield  {author} {\bibinfo {author} {\bibfnamefont {H.}~\bibnamefont
			{Yamaguchi}}, \bibinfo {author} {\bibfnamefont {T.}~\bibnamefont
			{Hashimoto}}, \bibinfo {author} {\bibfnamefont {S.}~\bibnamefont {Hayakawa}},
		\bibinfo {author} {\bibfnamefont {D.~N.}\ \bibnamefont {Binh}},  \emph
		{et~al.},\ }\href@noop {} {\bibfield  {journal} {\bibinfo  {journal} {Phys.
				Rev. C}\ }\textbf {\bibinfo {volume} {83}},\ \bibinfo {pages} {034306}
		(\bibinfo {year} {2011})}\BibitemShut {NoStop}%
	\bibitem [{\citenamefont {Volkov}(1965)}]{volkov_equilibrium_1965}%
	\BibitemOpen
	\bibfield  {author} {\bibinfo {author} {\bibfnamefont {A.}~\bibnamefont
			{Volkov}},\ }\href@noop {} {\bibfield  {journal} {\bibinfo  {journal} {Nucl.
				Phys.}\ }\textbf {\bibinfo {volume} {74}},\ \bibinfo {pages} {33} (\bibinfo
		{year} {1965})}\BibitemShut {NoStop}%
	\bibitem [{\citenamefont {Yamaguchi}\ \emph {et~al.}(1979)\citenamefont
		{Yamaguchi}, \citenamefont {Kasahara}, \citenamefont {Nagata},\ and\
		\citenamefont {Akaishi}}]{yamaguchi_effective_1979}%
	\BibitemOpen
	\bibfield  {author} {\bibinfo {author} {\bibfnamefont {N.}~\bibnamefont
			{Yamaguchi}}, \bibinfo {author} {\bibfnamefont {T.}~\bibnamefont {Kasahara}},
		\bibinfo {author} {\bibfnamefont {S.}~\bibnamefont {Nagata}}, \ and\ \bibinfo
		{author} {\bibfnamefont {Y.}~\bibnamefont {Akaishi}},\ }\href@noop {}
	{\bibfield  {journal} {\bibinfo  {journal} {Prog. Theor. Phys.}\ }\textbf
		{\bibinfo {volume} {62}},\ \bibinfo {pages} {1018} (\bibinfo {year}
		{1979})}\BibitemShut {NoStop}%
	\bibitem [{\citenamefont {Okabe}\ and\ \citenamefont
		{Abe}(1979)}]{okabe_structure_1979}%
	\BibitemOpen
	\bibfield  {author} {\bibinfo {author} {\bibfnamefont {S.}~\bibnamefont
			{Okabe}}\ and\ \bibinfo {author} {\bibfnamefont {Y.}~\bibnamefont {Abe}},\
	}\href@noop {} {\bibfield  {journal} {\bibinfo  {journal} {Prog. Theor.
				Phys.}\ }\textbf {\bibinfo {volume} {61}},\ \bibinfo {pages} {1049} (\bibinfo
		{year} {1979})}\BibitemShut {NoStop}%
	\bibitem [{\citenamefont {Brink}(1966)}]{brink_alpha-particle_1966}%
	\BibitemOpen
	\bibfield  {author} {\bibinfo {author} {\bibfnamefont {D.}~\bibnamefont
			{Brink}},\ }\href@noop {} {\emph {\bibinfo {title} {The Alpha-Particle Model
				of Light Nuclei}}},\ in {International} {School} of {Physics} ``{Enrico}
	{Fermi}", {Course} 37\ (\bibinfo {address} {in International School of
		Physics},\ \bibinfo {year} {1966})\BibitemShut {NoStop}%
	\bibitem [{\citenamefont {Ring}\ and\ \citenamefont
		{Schuck}(2004)}]{ring_nuclear_2004}%
	\BibitemOpen
	\bibfield  {author} {\bibinfo {author} {\bibfnamefont {P.}~\bibnamefont
			{Ring}}\ and\ \bibinfo {author} {\bibfnamefont {P.}~\bibnamefont {Schuck}},\
	}\href@noop {} {\emph {\bibinfo {title} {The Nuclear Many-Body Problem}}}\
	(\bibinfo  {publisher} {Springer Science \& Business Media},\ \bibinfo {year}
	{2004})\BibitemShut {NoStop}%
	\bibitem [{\citenamefont {Yamada}\ \emph {et~al.}(2008)\citenamefont {Yamada},
		\citenamefont {Funaki}, \citenamefont {Horiuchi}, \citenamefont {Ikeda} \emph
		{et~al.}}]{yamada_monopole_2008}%
	\BibitemOpen
	\bibfield  {author} {\bibinfo {author} {\bibfnamefont {T.}~\bibnamefont
			{Yamada}}, \bibinfo {author} {\bibfnamefont {Y.}~\bibnamefont {Funaki}},
		\bibinfo {author} {\bibfnamefont {H.}~\bibnamefont {Horiuchi}}, \bibinfo
		{author} {\bibfnamefont {K.}~\bibnamefont {Ikeda}},  \emph {et~al.},\
	}\href@noop {} {\bibfield  {journal} {\bibinfo  {journal} {Prog. Theor.
				Phys.}\ }\textbf {\bibinfo {volume} {120}},\ \bibinfo {pages} {1139}
		(\bibinfo {year} {2008})}\BibitemShut {NoStop}%
	\bibitem [{\citenamefont {Yamada}\ \emph {et~al.}(2012)\citenamefont {Yamada},
		\citenamefont {Funaki}, \citenamefont {Myo}, \citenamefont {Horiuchi} \emph
		{et~al.}}]{yamada_isoscalar_2012}%
	\BibitemOpen
	\bibfield  {author} {\bibinfo {author} {\bibfnamefont {T.}~\bibnamefont
			{Yamada}}, \bibinfo {author} {\bibfnamefont {Y.}~\bibnamefont {Funaki}},
		\bibinfo {author} {\bibfnamefont {T.}~\bibnamefont {Myo}}, \bibinfo {author}
		{\bibfnamefont {H.}~\bibnamefont {Horiuchi}},  \emph {et~al.},\ }\href@noop
	{} {\bibfield  {journal} {\bibinfo  {journal} {Phys. Rev. C}\ }\textbf
		{\bibinfo {volume} {85}},\ \bibinfo {pages} {034315} (\bibinfo {year}
		{2012})}\BibitemShut {NoStop}%
	\bibitem [{\citenamefont {Chiba}\ \emph {et~al.}(2016)\citenamefont {Chiba},
		\citenamefont {Kimura},\ and\ \citenamefont
		{Taniguchi}}]{chiba_isoscalar_2016}%
	\BibitemOpen
	\bibfield  {author} {\bibinfo {author} {\bibfnamefont {Y.}~\bibnamefont
			{Chiba}}, \bibinfo {author} {\bibfnamefont {M.}~\bibnamefont {Kimura}}, \
		and\ \bibinfo {author} {\bibfnamefont {Y.}~\bibnamefont {Taniguchi}},\
	}\href@noop {} {\bibfield  {journal} {\bibinfo  {journal} {Phys. Rev. C}\
		}\textbf {\bibinfo {volume} {93}},\ \bibinfo {pages} {034319} (\bibinfo
		{year} {2016})}\BibitemShut {NoStop}%
	\bibitem [{\citenamefont {Kanada-En'yo}(2016)}]{kanada-enyo_isovector_2016}%
	\BibitemOpen
	\bibfield  {author} {\bibinfo {author} {\bibfnamefont {Y.}~\bibnamefont
			{Kanada-En'yo}},\ }\href@noop {} {\bibfield  {journal} {\bibinfo  {journal}
			{Phys. Rev. C}\ }\textbf {\bibinfo {volume} {93}} (\bibinfo {year}
		{2016})}\BibitemShut {NoStop}%
	\bibitem [{\citenamefont {Chiba}\ \emph {et~al.}(2017)\citenamefont {Chiba},
		\citenamefont {Taniguchi},\ and\ \citenamefont
		{Kimura}}]{chiba_inversion_2017}%
	\BibitemOpen
	\bibfield  {author} {\bibinfo {author} {\bibfnamefont {Y.}~\bibnamefont
			{Chiba}}, \bibinfo {author} {\bibfnamefont {Y.}~\bibnamefont {Taniguchi}}, \
		and\ \bibinfo {author} {\bibfnamefont {M.}~\bibnamefont {Kimura}},\
	}\href@noop {} {\bibfield  {journal} {\bibinfo  {journal} {Phys. Rev. C}\
		}\textbf {\bibinfo {volume} {95}},\ \bibinfo {pages} {044328} (\bibinfo
		{year} {2017})}\BibitemShut {NoStop}%
	\bibitem [{\citenamefont {Kelley}\ \emph {et~al.}(2012)\citenamefont {Kelley},
		\citenamefont {Kwan}, \citenamefont {Purcell}, \citenamefont {Sheu} \emph
		{et~al.}}]{kelley_energy_2012}%
	\BibitemOpen
	\bibfield  {author} {\bibinfo {author} {\bibfnamefont {J.~H.}\ \bibnamefont
			{Kelley}}, \bibinfo {author} {\bibfnamefont {E.}~\bibnamefont {Kwan}},
		\bibinfo {author} {\bibfnamefont {J.~E.}\ \bibnamefont {Purcell}}, \bibinfo
		{author} {\bibfnamefont {C.~G.}\ \bibnamefont {Sheu}},  \emph {et~al.},\
	}\href@noop {} {\bibfield  {journal} {\bibinfo  {journal} {Nucl. Phys. A}\
		}\textbf {\bibinfo {volume} {880}},\ \bibinfo {pages} {88} (\bibinfo {year}
		{2012})}\BibitemShut {NoStop}%
	\bibitem [{\citenamefont {Stone}(2005)}]{stone_table_2005}%
	\BibitemOpen
	\bibfield  {author} {\bibinfo {author} {\bibfnamefont {N.}~\bibnamefont
			{Stone}},\ }\href@noop {} {\bibfield  {journal} {\bibinfo  {journal} {At.
				Data Nucl. Data}\ }\textbf {\bibinfo {volume} {90}},\ \bibinfo {pages} {75}
		(\bibinfo {year} {2005})}\BibitemShut {NoStop}%
	\bibitem [{\citenamefont {Kawabata}\ \emph
		{et~al.}(2007{\natexlab{b}})\citenamefont {Kawabata}, \citenamefont
		{Akimune}, \citenamefont {Fujita}, \citenamefont {Fujita} \emph
		{et~al.}}]{kawabata_dilute_2007}%
	\BibitemOpen
	\bibfield  {author} {\bibinfo {author} {\bibfnamefont {T.}~\bibnamefont
			{Kawabata}}, \bibinfo {author} {\bibfnamefont {H.}~\bibnamefont {Akimune}},
		\bibinfo {author} {\bibfnamefont {H.}~\bibnamefont {Fujita}}, \bibinfo
		{author} {\bibfnamefont {Y.}~\bibnamefont {Fujita}},  \emph {et~al.},\
	}\href@noop {} {\bibfield  {journal} {\bibinfo  {journal} {Nucl. Phys. A}\
		}\textbf {\bibinfo {volume} {788}},\ \bibinfo {pages} {301} (\bibinfo {year}
		{2007}{\natexlab{b}})}\BibitemShut {NoStop}%
	\bibitem [{\citenamefont {Yamada}\ and\ \citenamefont
		{Funaki}(2012)}]{yamada_three-body_2012}%
	\BibitemOpen
	\bibfield  {author} {\bibinfo {author} {\bibfnamefont {T.}~\bibnamefont
			{Yamada}}\ and\ \bibinfo {author} {\bibfnamefont {Y.}~\bibnamefont
			{Funaki}},\ }\href@noop {} {\bibfield  {journal} {\bibinfo  {journal} {Prog.
				Theor. Phys. Suppl.}\ }\textbf {\bibinfo {volume} {196}},\ \bibinfo {pages}
		{388} (\bibinfo {year} {2012})}\BibitemShut {NoStop}%
\end{thebibliography}
\end{document}